# Earth-abundant and Non-toxic SiX (X=S, Se) Monolayers as Highly Efficient Thermoelectric Materials

Ji-Hui Yang,[‡,§] Qinghong Yuan,[‡] Huixiong Deng,[∥] Su-Huai Wei[†*] and Boris I. Yakobson[‡*]

[‡]Department of Materials Science and Nanoengineering, Rice University, Houston, Texas 77005, USA

[∥]State Key Laboratory for Superlattices and Microstructures, Institute of Semiconductors, Chinese Academy of Sciences, P. O. Box 912, Beijing 100083, China

[§]National Renewable Energy Laboratory, Golden, CO 80401, USA

[†]Beijing Computational Science Research Center, Beijing 100094, China

**ABSTRACT:** Current thermoelectric (TE) materials often have low performance or contain less abundant and/or toxic elements, thus limiting their large-scale applications. Therefore, new TE materials with high efficiency and low cost are strongly desirable. Here we demonstrate that, SiS and SiSe monolayers made from non-toxic and earth-abundant elements intrinsically have low thermal conductivities arising from their low-frequency optical phonon branches with large overlaps with acoustic phonon modes, which is similar to the state-of-the-art experimentally demonstrated material SnSe with a layered structure. Together with high thermal power factors due to their two-dimensional nature, they show promising TE performances with large figure of merit (ZT) values exceeding 1 or 2 over a wide range of temperatures. We establish some basic understanding of identifying layered materials with low thermal conductivities, which can guide and stimulate the search and study of other layered materials for TE applications.

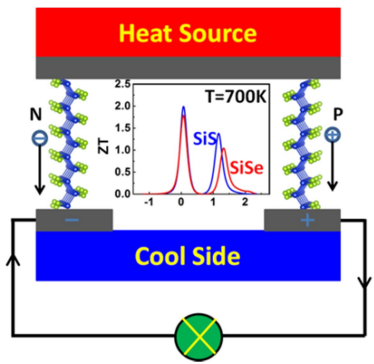

Thermoelectric materials have great potential for novel energy and environmental applications and have been extensively studied recently. The performance of a TE material is usually evaluated by the dimensionless figure of merit, which is denoted as ZT and defined as ZT = $(S^2\sigma/\kappa)T$, where S, σ, κ and T are the Seebeck coefficient, the electrical conductivity, the thermal conductivity (including both electronic contribution $\kappa_e$ and lattice contribution $\kappa_L$), and the average working temperature, respectively.[1-4] The product $S^2\sigma$ is also known as the power factor (PF). To be an ideal TE material with high heat-to-electric conversion efficiency, a high ZT value, i.e., larger than unity, is required.

Currently, two strategies are usually considered to enhance ZT of a TE material. One is to enhance the Seebeck coefficient or the PF using band structure engineering methods,[5] such as modifying the electronic density of states and Fermi energy through doping heteroatoms, reducing crystal symmetry to achieve high band edge degeneracy,[6-8] etc. The other is to reduce thermal conductivity using methods such as nanostructuring,[9-21] alloying,[14,22-26] etc. These strategies have achieved significant progress in the past decade. However, most of now-existing TE materials, including PbTe, $Bi_2Te_3$ and complex structure based TE materials,[3] strongly rely on heavy elements and/or rare-earth metals, which often have low abundance or are toxic,[3] thus limiting the large-scale employment of TE materials. Consequently, it will be of great importance to explore new TE materials not only with a high maximum ZT over a wide range of temperatures, but also made from non-toxic and earth-abundant elements.

Recently, those materials with layered structures have shown such promise. As an experimentally demonstrated state-of-the-art thermoelectric material, SnSe has a simple layered structure similar to black phosphorus and only contains earth-abundant and non-toxic elements. It has been experimentally reported that SnSe layers can achieve a ZT of 2.6 at T=923 K[27] and a ZT of 2.0 at T=773 K.[28] Such excellent TE performance is attributed to its intrinsically ultralow thermal conductivity from its strong anharmonicity and its ultrahigh power factor from its high electrical conductivity and strongly enhanced Seebeck coefficient enabled by the contribution of multiple electronic valence bands. When SnSe is further thinned to a monolayer, it can even achieve an enhanced ZT value of 3.27 at T=700 K by theoretical prediction.[29] Such enhanced TE performance in monolayer systems could partially be attributed to the sharp peaks in the electronic density of states (DOS) near the Fermi energy, which can significantly improve the Seebeck coefficient.[4,29] Attractive TE performance is also reported in monolayer phosphorene with a ZT value of 2.5 at T= 500 K.[30] Due to the attractive thermoelectric properties of layered materials, one would wonder what kind of layered materials can have promising thermoelectric performances and whether there are some clues to identify layered materials with low thermal conductivities and high electric conductivities and thus high thermoelectric performances.

Here, using first-principles calculation methods, we present that SiS and SiSe monolayers in their groud *Pmma* structure also show attractive properties for TE applications. These two systems intrinsically have not only low thermal conductivities but also high thermal power factors, which are similar to the case of SnSe with a layered structure. Their low thermal conductivities can be understood from their low-frequency optical phonon branches with large overlaps with acoustic phonon modes and their high thermal power factors are owing to their two-dimensional nature which can result in sharp peaks of density of states near the Fermi energy. Consequently, high ZT values of 1.06 (0.92) and 1.99 (1.79) can be achieved for SiS (SiSe), at T=500 K and T=700 K, respectively, with a conservative choice of the carrier relaxation time. Actually, larger ZT values surpassing 2 or 3 are expected because carrier relaxation time in monolayer materials is often large, which is justified in this work. With additional consideration that SiS and SiSe only consist of earth-abundant and non-toxic elements, they could be good candidates for high efficiency and large-scale TE applications. With a comparable study of the thermal conductivities of monolayer SiS(Se), SnSe, and phosphorene, we establish some basic understanding of identifying layered materials with low thermal conductivities, which is expected to guide and stimulate future search and study of other layered materials for TE applications.

The ground state structure of SiX monolayer is theoretically predicted in Reference 31 with *Pmma* symmetry and shown in Figure 1a. It can be seen as Si zigzag chains connected and sandwiched by two sulfur (selenium) layers. A*b-initio* molecular dynamics (AIMD) simulations (see Methods) are performed and indicate that SiS and SiSe monolayers can be stable at T up to 1000 K (Figures S1 and S2, Supporting Information), which guarantee their high-temperature TE applications. In case AIMD overestimates the stable temperature region, a medium–high temperature range (300─700 K) is considered for the study of their TE performance. Figures 1b and 1c show the band structures of SiS and SiSe with bandgaps corrected by using Heyd-Scuseria-Ernzerhof (HSE06) hybrid functional.[32] As can be seen, the Si zigzag (Γ─Y) direction is more disperse than Si-S(Se)-Si (Γ—X) direction with smaller effective masses for both holes and electrons (Supporting Information). As a result, it is expected from both the crystal structure and electronic structure that SiS and SiSe monolayers could have anisotropic thermal and electrical transport behaviors, which require separate consideration in calculating TE properties along these two anisotropic directions.

First, we discuss their thermal transport properties. At room temperature or high, the lattice thermal conductivity is mainly governed by the anharmonic phonon-phonon scattering and can be expressed as:[33-35]

$$\kappa_\text{L} = \frac{1}{NV_0}\sum_\lambda C_\lambda v_\lambda^2 \tau_\lambda, \qquad (1)$$

where $N$ is the total number of q points sampling the Brillouin zone, $V_0$ is the volume of the primitive cell, $\lambda$ is the phonon mode index including both band and q point, $C_\lambda$ is the specific heat, $v_\lambda$ is phonon velocity, and $\tau_\lambda$ is phonon lifetime. To get accurate values of thermal conductivities, both acoustic contributions and contributions of optical phonon braches are taken into account with all the mode-dependent $C_\lambda$, $v_\lambda$ and $\tau_\lambda$ calculated using second-order and third-order force constants (FCs), which can be obtained from finite displacement method implemented in phono3py.[36] Details of the calculations can be found in the Methods part. This method has achieved good agreement between theoretical calculations and experiments for many bulk systems.[36] To make sure the method is also reliable in monolayer systems, lattice thermal conductivities of SiS, SiSe, black phosphorus (BP), and SnSe monolayers are comparably investigated. Our calculated thermal conductivities of BP and SnSe monolayers (see Table I) agree very well with previous calculations[29,35] using ShengBTE code, [33] in which mode-dependent phonon lifetimes are similarly calculated from the anharmonic third-order FCs.

Figures 2a and 2c show our calculated thermal conductivities of SiS and SiSe monolayers along two anisotropic directions at different temperatures. Compared to those of BP and SnSe monolayers (Figure S3, Supporting Information), the thermal conductivities of SiS and SiSe monolayers are about 1 order lower than that of phosphorene and only 2~3 times larger than that of SnSe monolayer. Besides, both SiS and SiSe show relatively weak anisotropic $\kappa_\text{L}$ similar to SnSe: in SiS, $\kappa_\text{L}$ along Si zigzag chains is slightly smaller and in SiSe, Si-Se-Si direction has a slightly smaller $\kappa_\text{L}$ (see explanation in the Supporting Information). This kind of unexpected weak anisotropy might result from their low-frequency optical modes, as discussed below.

To understand the relatively low $\kappa_L$ in SiS and SiSe, their phonon dispersion is analyzed with the effect of nonanalytical term correction considered (Figures 2b and 2d). Compared to the phonon dispersion of BP and SnSe monolayers (Figure S3, Supporting Information), we find that: (1) SiS and SiSe generally have smaller acoustic phonon velocities than BP (see also Table I); (2) The optical branches of SiS and SiSe have low frequencies and some of them overlap with the acoustic branches, similar to SnSe (Figure S3, Supporting Information). These phonon dispersion characters can be seen as indication of soft modes and strong anharmonicity, which is further confirmed by the estimation of mode Grüneisen parameters (Figure S4, Supporting Information). Consequently, low thermal conductivities are obtained. We note that, the low-frequency optical modes can contribute more than 30% to the total thermal conductivity in SnSe[29] and thus should be included in the calculations to avoid the underestimation of $\kappa_L$, as we did in this work. Such high contributions of optical modes might be the reason of weak anisotropic thermal transport properties in SiS, SiSe, and SnSe monolayers.

The electrical transport properties of SiS and SiSe monolayers are investigated by solving semi-classical Boltzmann transport equations using the method implemented in BoltzWann code[37] with the constant relaxation time approximation, which is often adopted in first-principles TE calculations.[38-40] Our calculated results for the electrical transport properties of SiS and SiSe monolayers at three different temperatures (T=300 K, 500 K, and 700 K) are shown in Figure 3. Note that, except the Seebeck coefficient S, the electrical conductivity $\sigma$, the power factor $S^2\sigma$, and the electrical thermal conductivity $\kappa_e$ are all dependent on the choice of the carrier relaxation time constant $\tau$, which in principle is band and temperature dependent.[37,41] In practice, $\tau$ is often determined by fitting the experimental values.[29,37,42] Alternatively, $\tau$ can also be systematically determined from first-principles calculations, by either considering explicitly the electron-phonon scattering mechanisms[41] or using the deformation potential approximation for mobility calculations.[30] Under this approximation, the carrier mobility $\mu$ can be calculated as:[43,44]

$$\mu = \frac{e\hbar^3 C}{k_B T m_e^* m_d (E_1^i)^2} , (2)$$

where $m_e^*$ is the carrier effective mass along the transport direction, $m_d$ is the carrier average effective mass, $E_1^i$ represents the deformation potential constant of the valance band maximum (VBM) for hole or conduction band minimum (CBM) for electron along the transport direction and $C$ is the elastic modulus of the longitudinal strain in the propagation directions of the longitudinal acoustic wave (see more details in the Supporting Information). This expression is widely used to estimate the carrier mobility in monolayer systems and good agreement between first-principles calculations and experiments are achieved.[30,,43,44] Once the mobility $\mu$ is known, electric conductivity can be obtained from $\sigma = en\mu$, where $n$ is the carrier density and $e$ is the elementary charge. Consequently, $\tau$ can be systematically determined from Boltzmann transport equations and Equation (2).

Using the above method, we obtain $\tau$ for both hole and electron transport along Si zigzag and Si-S(Se)-Si directions in a wide range of temperatures (Figure S6, Supporting Information). The relaxation time decreases with the increase of temperatures as expected. What's more important, $\tau$ is larger than 40 femtoseconds (*fs*) in general for both hole and electron transportations along both directions at T=300 K and larger than 20 *fs* at T=700 K. For some transportation of specific carriers along certain directions, $\tau$ can be even larger than hundreds of *fs*. Notably, in the recent TE work on SnSe monolayer,[29] $\tau$ is fitted to be about 30 *fs* using the experimental value of electrical conductivity at T=300 K. Similarly, in the case of transition metal dichalcogenides such as WS$_2$, $\tau$ is fitted to be about 37 *fs*.[42] In the case of phosphorene, $\tau$ can be as large as about 150 *fs* using the theoretical carrier mobility at T=300 K or about 70 *fs* using the experimental value of mobility at room temperature.[30] The highly increased $\tau$ in monolayer systems compared to that in bulk TE materials could be attributed to the quantum confinement effects, which can reduce electron-phonon scattering by discretizing the electronic spectrum with energy differences that do not match phonon frequencies.[41,45] This is another indication that monolayer materials could be better for potential TE applications. Compared to $\tau$ used in previous low-dimensional works[29,30,42] and considering 10 *fs* is often used in bulk systems, 20 *fs* is the most conservative choice of $\tau$ for SiS and SiSe monolayers.

Figure 3 shows the electrical transport properties of SiS and SiSe monolayers. As seen in Figure 3b, SiS monolayer has a Seebeck coefficient maximum of about 0.002 V K$^{-1}$ at T=300 K for both *p*-type and *n*-type along both directions. This value is similar to that of SnSe[29] and BP monolayers and several times higher than typical values of bulk TE materials.[29] Such high Seebeck coefficient can be attributed to quantum confinement effect which induces sharp DOS peaks near the Fermi energy. The power factor of SiS is also comparable to that of SnSe monolayer[29] with peak values of 0.016 W K$^{-2}$m$^{-2}$ for *p*-type and 0.011 W K$^{-2}$m$^{-2}$ for *n*-type along Si zigzag direction with τ=20 *fs*. While the Seebeck coefficient of SiS show very weak anisotropy, the power factors along Si zigzag direction is much larger than those along Si-S-Si direction. This can be ascribed to the much smaller carrier effective masses in this Si zigzag direction (Supporting Information). Besides, due to larger DOS near the VBM than that near the CBM,[31] thermal power factors for *p*-type cases are much larger than those for *n*-type cases along Si zigzag direction. SiSe generally have similar electrical transport properties, as shown in Figure 3.

Now that we have both lattice thermal conductivities and electrical transport properties, ZT values of SiS and SiSe can be obtained using the formula $ZT = (S^2\sigma/\kappa)T$. Figures 3d and 3g show our calculated ZT of SiS and SiSe with τ of 20 *fs*. As we can see, ZT along Si zigzag directions always has larger values than that along Si-S(Se)-Si direction. This is because the thermal power factors have much stronger anisotropies than lattice thermal conductivities, thus dominating the anisotropies of ZT. In addition, the thermal power factors along Si zigzag direction have larger values for *p*-type cases. Therefore, ZT for *p*-type SiS and SiSe are always larger than that for *n*-type cases.

Our calculations show that SiS (SiSe) monolayer can achieve ZT peak values of 1.06 (0.92) and 1.99 (1.79) along Si zigzag direction for *p*-type cases at T=500 K and T=700 K, respectively, with a conservative choice of carrier relaxation time of 20 *fs*. For *n*-type cases, ZT values of SiS (SiSe) monolayer can achieve 0.67 (0.48) and 1.38(1.04) at T=500 K and T=700 K, respectively. Note that, ZT values increase with larger τ, as shown in Figure S7. For example, if τ is chosen as 50 *fs*, ZT peak values can surpass 3. The choice of 20 *fs*, as discussed in the above discussions, gives the most conservative estimation of ZT values for SiS and SiSe monolayers. Even in this case, the large ZT values make these two monolayer materials very attractive for TE applications, especially when taking into consideration that both SiS and SiSe can be made from earth-abundant and non-toxic elements. ZT values can be further enhanced by alloying these two materials together to reduce their thermal conductivities. The electron chemical potential positions corresponding to ZT peak values are about 0.05~0.10 eV above the VBM or below the CBM of SiS or SiSe with carrier densities of about $10^{12}$~$10^{13}$ cm$^{-2}$, which is not difficult to achieve in term of doping.[46] In addition, SiS and SiSe monolayers can be vertically stacked to form layered bulks and their TE performance is expected to be little affected due to the weak interlayer interactions with large layer distances (Supporting Information).

In summary, we have performed systematic study of the thermoelectric properties of SiS and SiSe monolayers. We find that they have intrinsically low thermal conductivities due to low electron-phonon scattering and high thermal power factors due to high DOS near the band edge. Consequently, both SiS and SiSe are expected to be promising thermoelectric materials with attractive ZT values at medium high temperatures. The basic understanding established in this work could shed light on intrinsic thermoelectric properties of layered materials and provide guidence on future search and study of other layered materials for thermoelectric applications.

**METHODS**

*First-principles Calculation Methods.* Our first-principles calculations were performed using density-functional theory (DFT)[47,48] as implemented in the VASP code[49,50]. The electron and core interactions are included using the frozen-core projected augmented wave (PAW) approach.[51] For all the structure relaxations, we used generalized gradient approximation (GGA) formulated by Perdew, Burke, and Ernzerhof (PBE).[52] The structures are relaxed until the atomic forces are less than 0.001 eV/Å and total energies are converged to $10^{-8}$ eV with the cutoff energy for plane-wave basis functions set to 400 eV. We used HSE06 hybrid functional with default parameters to correct bandgap errors.[32]

*AIMD Simultions.* To examine the thermal stabilities of SiS and SiSe at the studied temperature range, AIMD simulations are performed using $3 \times 5 \times 1$ supercells containing 120 atoms with single Gamma point. We adopted NVT canonical ensembles and the time step is set in the range of 0.3~2.5 fs. The total simulation time is more than 10 *ps* to make sure the conclusions are reliable.

*Calculation of lattice thermal conductivities.* The lattice thermal conductivity is calculated using the methods implemented in Phono3py.[36] Finite displacement method is used to calculate the force constants with the help of first-principles calculations. The second-order force constants of SiS and SiSe are calculated using $3 \times 5 \times 1$ supercells containing 120 atoms with k-point meshes of $2 \times 2 \times 1$. The third-order force constants of SiS and SiSe are calculated using $2 \times 3 \times 1$ supercells containing 48 atoms with k-point meshes of $2 \times 2 \times 1$. With symmetry consideration, the total number of displacements is 3570 in this case. After obtaining the force constants, a dense k-point mesh of $10 \times 15 \times 1$ is used to calculate the lattice thermal conductivities of SiS and SiSe to make sure the final results are well converged. To calculate BP and SnSe monolayers, $6 \times 6 \times 1$ supercells with 144 atoms are used to get the second-order force constants and $3 \times 3 \times 1$ supercells with 36 atoms are used to get the third-order force constants. The total numbers of displacements for BP and SnSe are 701 and 1402 respectively. After obtaining the force constants, a dense k-point mesh of $30 \times 30 \times 1$ is used to calculate the lattice thermal conductivities of BP and SnSe monolayers. Note that, the calculated thermal conductivities should be normalized by multiplying $L_z/d$, where $L_z$ is the length along z (vacuum) directions and d is the thickness of the layers. Here we used bulk layer distances assuming SiS and SiSe bulks are AA-stacking. Similar normalizations are done for BP and SnSe monolayers.

## SUPPORTING INFORMATION

Detailed structural information of SiS and SiSe monolayers, AIMD simulation results of SiS and SiSe, calculated lattice thermal conductivities and phonon dispersion of BP and SnSe monolayers for comparative study, explanations of weak anisotropy of $\kappa_L$ in SiS and SiSe, calculated mode Grüneisen parameters of SiS, SiSe, BP and SnSe monolayers, mobility calculations and effective masses of SiS and SiSe, both hole and electron relaxation-time dependence of temperatures along two anisotropic directions of SiS and SiSe, and ZT dependence on the choice of τ. This material is available free of charge via the Internet at http://pubs.acs.org.

## AUTHOR INFORMATION
## CORRESPONDING AUTHOR
* suhuaiwei@csrc.ac.cn; biy@rice.edu
**Notes**
The authors declare no competing financial interest.

## ACKNOWLEDGMENT
Work at Rice was supported by the U.S. Army Research Office MURI Grant W911NF-11-1-0362. Effort at NREL is funded by the Laboratory Directed Research and Development Program under Grant No. 065K1601. The work at Institute of Semiconductors, Chinese Academy of Sciences was supported by National Natural Science Foundation of China under Grants No. 11474273. The calculations are done on Rice DAVinCI super-computer and NREL peregrine super-computer.**REFERENCES**

1. Heremans, J. P.; Dresselhaus, M. S.; Bell, L. E.; Morelli, D. T. When Thermoelectrics Reached the Nanoscale. *Nature Nanotechnol.* **2013**, 8, 471–473.

2. Zhao, L. D.; Dravid, V. P.; Kanatzidis, M. G. The Panoscopic Approach to High Performance Thermoelectrics. *Energ. Environ. Sci.* **2014**, 7, 251–268.

3. Snyder G. J.; Toberer, E. S. Complex Thermoelectric Materials. *Nature Mater.* **2008**, 7, 105-114.

4. Kovalenko, M. V.; Manna, L.; Cabot, A.; Hens, Z.; Talapin, D. V.; Kagan, C. R.; Klimov, V. I.; Rogach, A. L.; Reiss, P.; Milliron, D. J.; Guyot-Sionnnest, P.; Konstantatos, G.; Parak, W. J.; Hyeon, T.; Korgel, B. A.; Murray, C. B.; Heiss, W. Prospects of Nanoscience with Nanocrystals. *ACS Nano* **2015**, *9*, 1012–1057.

5. Pei, Y.; Wang, H.; Snyder, G. J. Band Engineering of Thermoelectric Materials. *Adv. Mater.* **2012**, *24*, 6125–6135.

6. Heremans, J. P.; Jovovic, V.; Toberer, E. S.; Saramat, A.; Kurosaki, K.; Charoenphakdee, A.; Yamanaka S.; Snyder, G. J. Enhancement of Thermoelectric Efficiency in PbTe by Distortion of the Electronic Density of States. *Science* **2008**, *321*, 554–557.

7. Liu, W.-S.; Zhao, L.-D.; Zhang, B.-P.; Zhang H.-L.; Li, J.-F. Enhanced Thermoelectric Property Originating from Additional Carrier Pocket in Skutterudite Compounds. *Appl. Phys. Lett.* **2008**, *93*, 042109.

8. Pei, Y.; Shi, X.; LaLonde, A.; Wang, H.; Chen L.; Snyder, G. J. Convergence of Electronic Bands for High Performance Bulk Thermoelectrics. *Nature* **2011**, *473*, 66–69.

9. Zhang, Y.; Wang, H.; Kraemer, S.; Shi, Y.; Zhang, F.; Snedaker, M.; Ding, K.; Moskovits, M.; Snyder, G. J.; Stucky, G. D. Surfactant-Free Synthesis of $Bi_2Te_3$-Te Micro-Nano Heterostructure with Enhanced Thermoelectric Figure of Merit. *ACS Nano* **2011**, *5*, 3158–3165.10

Table I. Calculated phonon velocities and lattice thermal conductivities of SiS, SiSe, BP, and SnSe monolayers along two anisotropic directions. The armchair directions or Si-S(Se)-Si directions are denoted as 1 and the zigzag directions are denoted as 2. The velocity unit is m s$^{-1}$ and the thermal conductivities are given at T=300 K with unit of W m$^{-1}$K$^{-1}$. Available data in references are also given for comparison.

| system | $v_{1,LA}$ | $v_{1,TA}$ | $v_{2,LA}$ | $v_{2,TA}$ | $\kappa_{L,1}$ | $\kappa_{L,2}$ |
|---|---|---|---|---|---|---|
| SiS | 3319 | 7522 | 3714 | 2533 | 7.01 | 6.42 |
| SiSe | 1787 | 5499 | 4063 | 1688 | 4.19 | 5.77 |
| BP | 4135 | 4053 | 8454 (8397)[30] | 3885 (4246)[30] | 31.06 (24.3)[35] | 74.72 (83.5)[35] |
| SnSe | 3225 (3545)[29] | 2143 (1906)[29] | 2528 | 3318 | 2.18 (2.02)[29] | 2.25 (2.5)[29] |

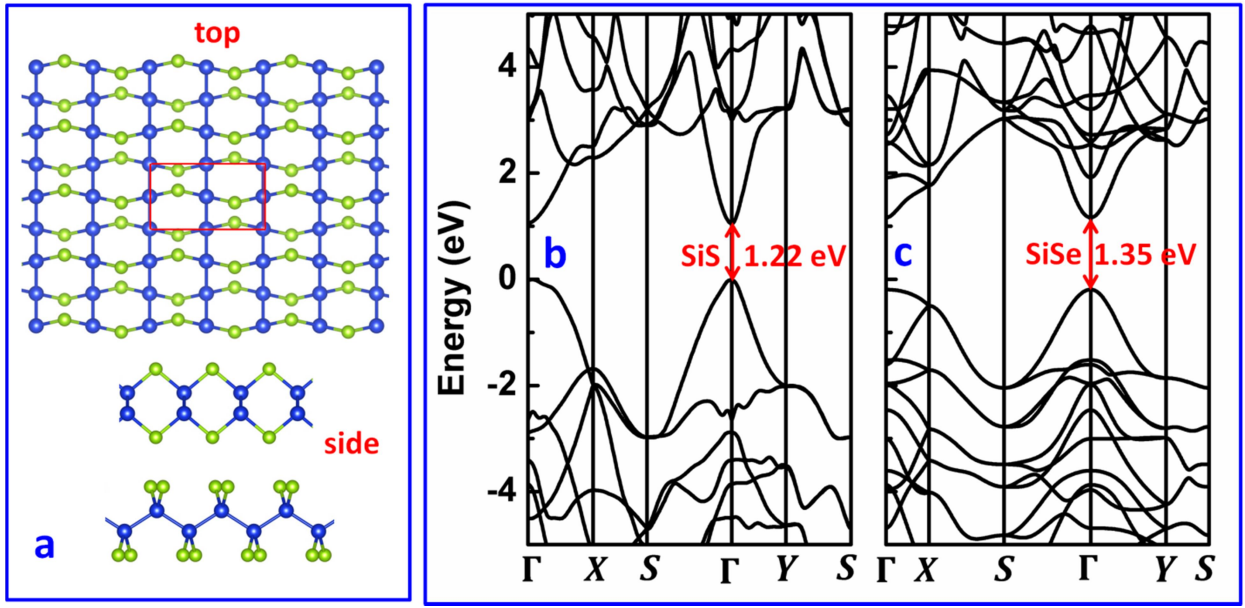

**Figure 1**. a) Top and side views of structure for *Pmma*-SiS(Se). Blue is for Si and green is for S or Se atoms. b) and c) HSE06 band structures of SiS and SiSe.

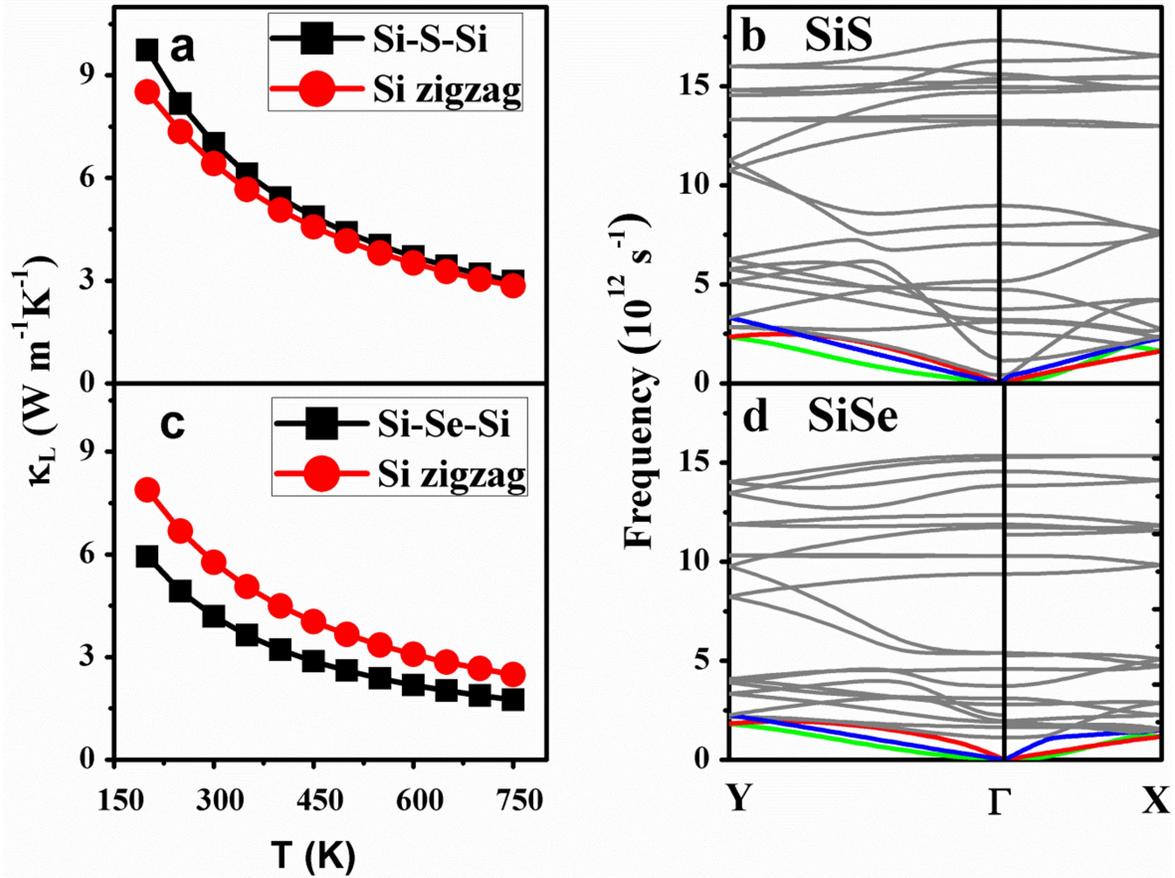

**Figure 2**. a) and c) Lattice thermal conductivity dependence on temperature for SiS and SiSe along Si zigzag chain and Si-S(Se)-Si directions. b) and d) Phonon dispersion for SiS and SiSe along Y─Γ (Si zigzag) and Γ─X [Si-S(Se)-Si] directions. Green lines represent ZA modes, blue lines represent TA modes, red lines represent LA modes, and grey lines represent optical modes.

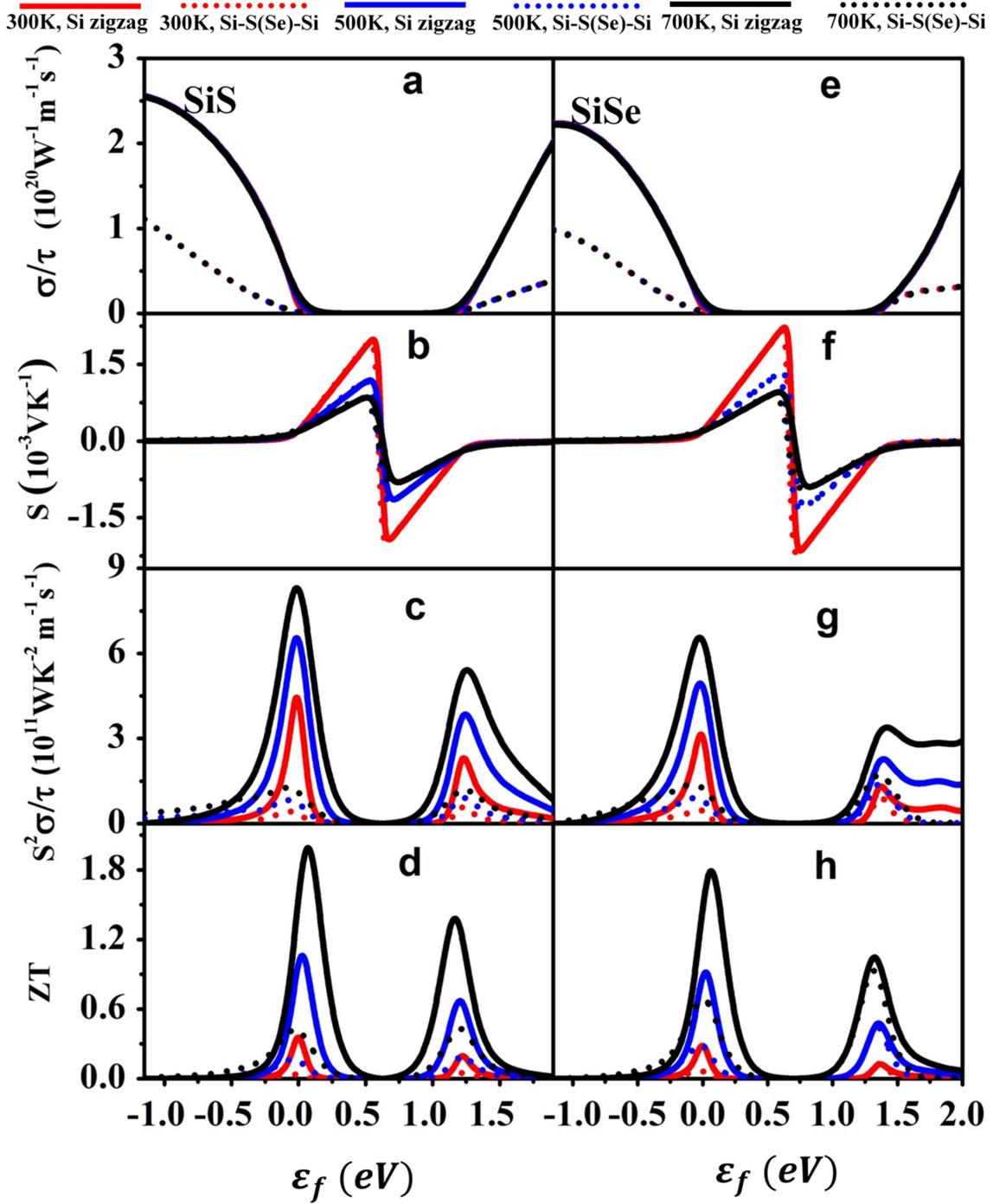

**Figure 3.** a) and e) Electrical conductivity, b) and f) Seebeck coefficient, c) and g) power factor, d) and h) ZT of SiS and SiSe, as functions of electron chemical potential $\varepsilon_f$ at three different temperatures T=300 K, 500 K, and 700 K and along two anisotropic directions, respectively. Here $\varepsilon_f$ is referenced to the VBM of SiS or SiSe.

18